\documentclass[a4paper,fleqn,usenatbib]{mnras}

\usepackage{newtxtext,newtxmath}

\usepackage[T1]{fontenc}
\usepackage{ae,aecompl}

\usepackage{siunitx}
\usepackage{natbib}
\usepackage{subfig}
\usepackage{gensymb}
\usepackage{floatrow}

\usepackage{graphicx}	
\usepackage{amsmath}	
\usepackage{amssymb}	

\title[Gaian Feedbacks \& Climate Perturbations]{Robustness of Gaian Feedbacks to Climate Perturbations}

\author[O. D. N. Alcabes et al.]{
Olivia D. N. Alcabes,\thanks{E-mail: oalcabes@uchicago.edu}
Stephanie Olson, and
Dorian S. Abbot 
\\
Department of Geophysical Sciences, The University of Chicago, Chicago, IL 60637
}

\date{Accepted for publication by MNRAS January 11th, 2020.}

\pubyear{2020}

\begin{document}
\label{firstpage}
\pagerange{\pageref{firstpage}--\pageref{lastpage}}
\maketitle

\begin{abstract}
The Gaia hypothesis postulates that life regulates its environment to be favorable for its own survival.
Most planets experience numerous perturbations throughout their lifetimes such as asteroid impacts, volcanism, and the evolution of 
their host star's luminosity.
For the Gaia hypothesis to be viable, life must be able to keep the conditions of its host planet habitable, even in the face of these challenges.
ExoGaia, a model created to investigate the Gaia hypothesis, has been previously used to demonstrate that a randomly mutating biosphere is in some cases capable of maintaining planetary habitability. However, those model scenarios assumed that all non-biological planetary parameters were static, neglecting the inevitable perturbations that real planets would experience.
To see how life responds to climate perturbations to its host planet,
we created three climate perturbations in ExoGaia: one rapid cooling of a planet and two heating events, one rapid and one gradual.
The planets on which Gaian feedbacks emerge
without climate perturbations are the same planets on which life is most likely to survive each of our perturbation scenarios.
Biospheres experiencing gradual changes to the environment are able to survive changes of larger magnitude than those experiencing rapid perturbations, and the magnitude of change matters more than the sign.
These findings suggest that if the Gaia hypothesis is correct, then typical perturbations that a planet would experience may be unlikely to disrupt Gaian systems.
\end{abstract}




\section{Introduction}

In the past few decades, there have been increasingly numerous detections of planets outside of our solar system \citep{fressin2013false,morton2014radius}. Although most of the planets that have been detected are too close to their host star to be habitable, detections of potentially habitable planets have been made and are expected to increase as detectors become more sensitive~\citep{Gillon:2017aa,Anglada-Escude:2016aa,Kreidberg:2014aa}. New initiatives such as the James Webb telescope will provide extensive, more detailed detections of exoplanets, and present the possibility of observing atmospheric spectra of potentially habitable planets~\citep{gardner2006james}.
These atmospheric spectra may reveal signs that life is present on some planets~\citep{DesMarais2002,schwieterman2018}.
\par
Gaian feedbacks may affect habitability and life on exoplanets. The Gaia hypothesis suggests that the biosphere of an inhabited planet has an active role in regulating the conditions on the planet such that the environment remains habitable~\citep{LOVELOCK1972579,lovelock1974,lovelock2000gaia}.
For example, one of the original applications of the Gaia hypothesis was the Faint Young Sun Paradox~\citep{lovelock1974}: 
even though the Sun's luminosity has increased by approximately 50\% of its initial value over the course of Earth's history, mean global temperatures have not changed dramatically. As a result, Earth has remained a habitable environment for billions of years~\citep{Sagan52},
and life on Earth has persisted uninterrupted on the planet since it emerged.
The Gaia hypothesis suggests that Earth remained habitable because Earth's biosphere regulated the climate as it was perturbed by the Sun.
\par
The Gaia hypothesis may also provide insight into the effects of rapid perturbations on a planet's climate and habitability.
Some examples of rapid climate perturbations throughout Earth's history are the Cretaceous-Paleogene mass extinction from the Chicxulub asteroid impact~\citep{Alvarez1980,Schulte1214} and "snowball Earth" glaciations.
Both perturbations cooled Earth's temperatures significantly, after which the climate recovered. This was at least in part due to the decrease in photosynthetic draw down of carbon dioxide in response to global cooling, allowing warming.
Looking to the future, anthropogenic changes to the environment may cause similarly significant climate perturbations~\citep{Lenton1066}.
Two such perturbations include global climate change~\citep{Archer2005} or the possibility of a nuclear winter~\citep{Turco1283}. It is unclear whether biological feedbacks would be able to stabilize Earth's climate during those perturbations.
Exoplanets are also likely to experience rapid climate perturbations, and long-term survival of life on exoplanets requires that these alien biospheres are as robust in the face of perturbations as life on Earth has been.
\par
There are multiple models that employ the Gaia hypothesis to qualitatively demonstrate how a biosphere could regulate the environment. Among them are Daisyworld, the Flask Model~\citep{Williams:2008aa} and Greenhouse World~\citep{worden2010notes}, which are precursors to the model used in this study, ExoGaia. ExoGaia is a conceptual model created by \citet{nicholson2018gaian} that investigates the effects of evolving microbial biospheres on the habitability of a planet.
ExoGaia assumes a simple climate with atmospheric gasses that either warm or cool the planet. Gasses can be created by both geochemical and biological processes, and biological functioning depends on the
planetary temperature.
Using ExoGaia, \citet{nicholson2018gaian} found that
planets are more likely to evolve Gaian feedbacks if they have a large number of links in their geochemical networks. However, \citet{nicholson2018gaian} focused on planets with constant climate forcing.
\par
Our goal in this paper is to consider the effects of perturbations, and to determine whether the same geochemical characteristics
that allow Gaian regulation without external perturbations also make life on a planet more likely to survive external perturbations.
We review the ExoGaia model and our modifications to it in Section \ref{sec:methods}. We give our main results in Section \ref{sec:results} and discuss them in Section \ref{sec:discussion}. We conclude in Section \ref{sec:conclusion}.
\par

\section{Methods} \label{sec:methods}

\subsection{Model Description}

Our study uses the ExoGaia model~\citep{nicholson2018gaian}.
ExoGaia models the interactions between two systems:
abiotic geochemical processes and a population of microbes.
Each planet is characterized by a specific chemical set and a time invariant set of randomized geological links between those chemicals.
Chemicals in ExoGaia do not correspond to real-world chemicals, and they have the following properties:

\begin{enumerate}
	\item they are gaseous
	\item they either warm or cool the planet, but never both
	\item they do not react with multiple chemicals at a time, i.e. $A \rightarrow B$, but $A + B \nrightarrow C$, for simplicity
	\item they have a constant inflow from an abiotic source (for example, background volcanism)
	\item they influence the planet exclusively through atmospheric abundance and their insulating or reflecting property
\end{enumerate}

The atmospheric composition directly affects the temperature of the planet.
\citet{nicholson2018gaian} defines variables $AI$ and $AR$, where $AI$ is the fraction of the planet's thermal energy that is retained through insulation, and $AR$ is the fraction of incoming radiation reflected. Both $AI$ and $AR$ are dependent on the amount of insulating or reflecting chemicals in the atmosphere~\citep{nicholson2018gaian}.
The radiative forcing variable, $\beta_{force}$, represents any radiative process affecting the planet's climate. The most obvious example of radiative forcing is the luminosity of a planet's host star. Other examples could include an asteroid impact, a volcanic eruption, or other changes in the planet's background greenhouse gas profile or aerosols. ExoGaia relates the temperature of the planet, $T$, to the amount of chemicals present and the radiative forcing using the following equation:
\\
\begin{equation}
	(1-AI)T = (1-AR) \beta_{force}
	\label{eq: climate}
\end{equation}
\\
In order for the temperature of the planet to remain stable, the insulating and reflecting chemicals must be altered in such a way that $\frac{(1-AR)}{(1-AI)}\beta_{force}$ is constant.
Temperature in ExoGaia is abstracted, so the unit for the planetary temperature does not correspond to Celsius, Kelvin, or any other familiar unit~\citep{nicholson2018gaian}.
\par

The biosphere in ExoGaia interacts with the
pre-existing
geochemical links. Each microbe belongs to a species, and each species consumes one chemical and expels a different chemical as waste.
Microbial metabolisms are affected by the temperature of the host planet, and their metabolisms function the best at their preferred temperature, $T_{pref}$~\citep{nicholson2018gaian}. In our experiment, all microbes have the same preferred temperature, $T_{pref} = 1000$, as in \citet{nicholson2018gaian}. Microbes are seeded on a planet while its atmosphere is still forming, once the temperature of the planet reaches the microbe's $T_{pref}$. If the preferred temperature is not reached, then the planet is seeded at the 50,000th timestep~\citep{nicholson2018gaian}.
The microbes are able to alter the temperature of the planet based on whether the chemicals they eat and excrete are reflecting or insulating. They have a temperature-dependent metabolism, so as the planet's temperature gets further away from their preferred temperature, they are more likely to die. \par
New microbes have a random chance of mutating, resulting in new species that
consume and expel different chemicals
 from the parent microbe. Therefore, the chemical links created by biology on the planet are able to change rapidly as new species evolve. In order to maintain a stable temperature as $\beta_{force}$ changes, microbes on a planet must be able to regulate the $AI$ and $AR$ variables
via their metabolisms~\citep{nicholson2018gaian}.
\par

\subsection{Experimental Procedures} \label{sec:experimental_procedures}
 We perform the following procedure for each radiative forcing experiment~\citep{nicholson2018gaian}:

\begin{enumerate}
	\item establish the planet's geochemical network by creating randomized invariant geological links between different chemicals
	\item plant a random microbe seeding on the planet when the planet reaches the preferred temperature
	of the microbes, which is equal to 1000 for all species
	\item run the simulation for \num{5e5} timesteps after seeding
	\item repeat steps (ii) and (iii) 10 times
	\item repeat steps (i)-(iv) 100 times
\end{enumerate}

The connectivity parameter establishes the likelihood that 
an abiotic link 
between two chemicals on a planet will form. We keep the connectivity parameter constant at C=0.4~\citep{nicholson2018gaian}.
We test 10 different biological seedings on each planet as opposed to 100, as we find that experiments with less seedings have similar qualitative results to experiments with more.
In order to categorize our results, we use a system very similar to that used by \citet{nicholson2018gaian}:\\
\\
\textbf{Abiding:} planets where all seedings survive\\
\textbf{Bottleneck:} planets where seedings either die early, before the 1000th timestep after seeding,
 or survive for the duration of the experiment\\
\textbf{Critical:} planets where some seedings survive, but others die out at random times\\
\textbf{Doomed:} planets where all seedings die\\
\par
 
Differing from \citet{nicholson2018gaian}, we change the definition of Critical slightly.
We require all Critical planets to have at least one surviving seeding by the end of the experiment, whereas \citet{nicholson2018gaian} still called planets Critical if they all died, so long as they died at random times.
We also call planets where all seedings die Doomed, as opposed to separating planets into Doomed and Extreme based on temperatures as \citet{nicholson2018gaian} did. 
\par
\begin{figure}
\centering
\captionsetup[subfigure]{width=0.45\columnwidth}
\sidesubfloat[]{
	\centering
	\includegraphics[width=0.4\columnwidth]{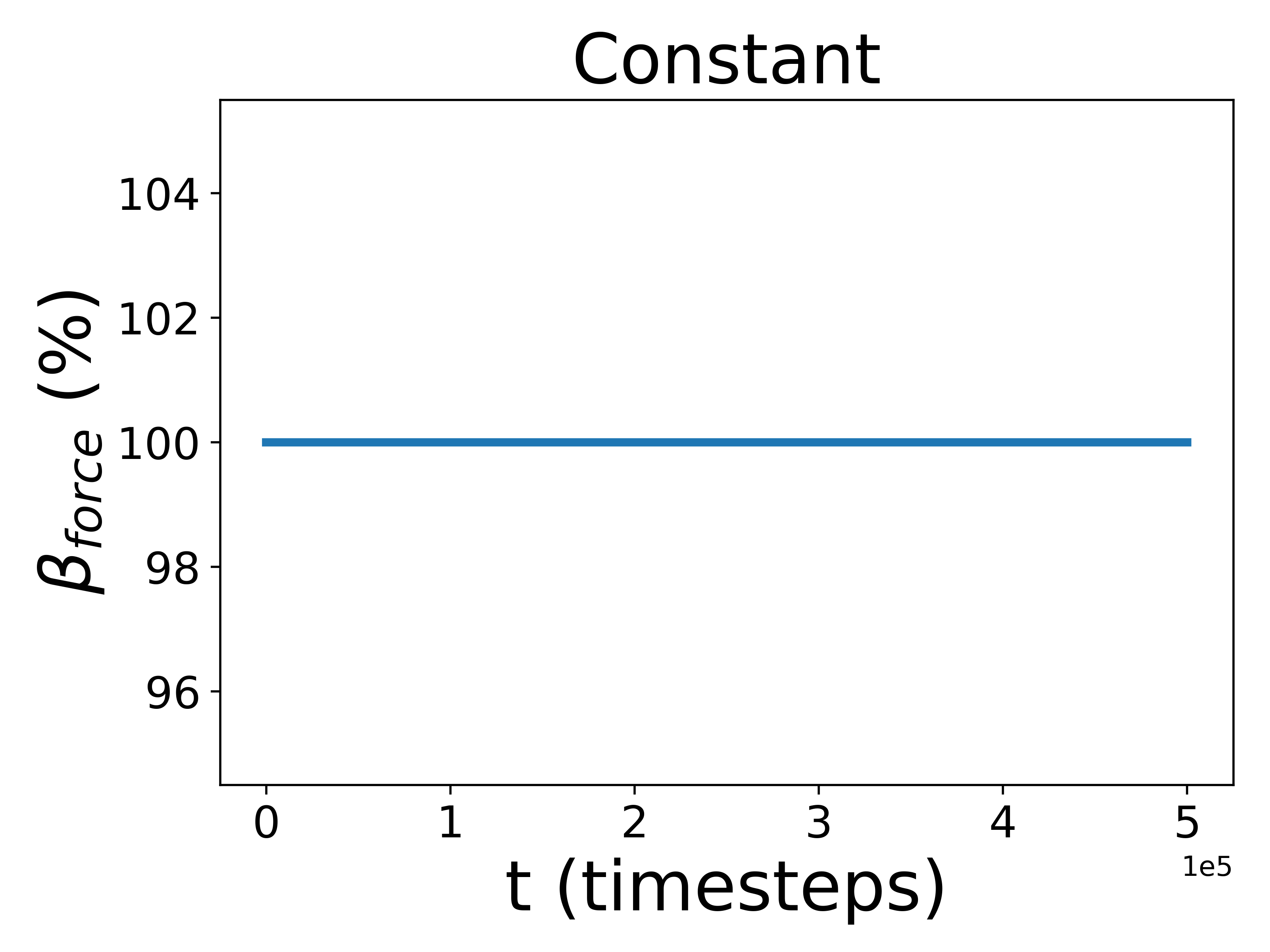}
	\label{fig:perturbation_control_example}
	}
\sidesubfloat[]{
	\centering
	\includegraphics[width=0.4\columnwidth]{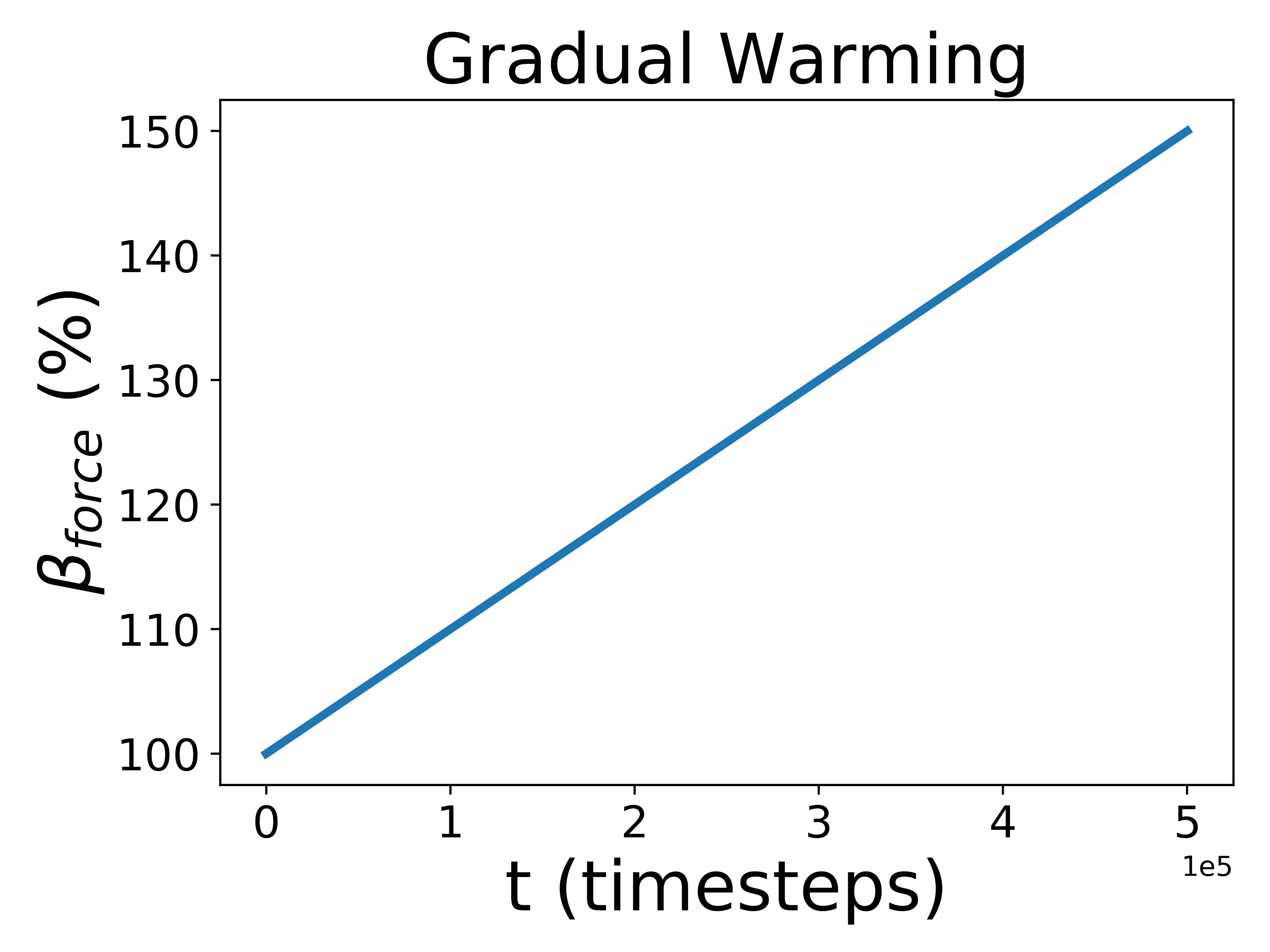}
	\label{fig:perturbation_gradual_example}
	} \\
\sidesubfloat[]{
	\centering
	\includegraphics[width=0.4\columnwidth]{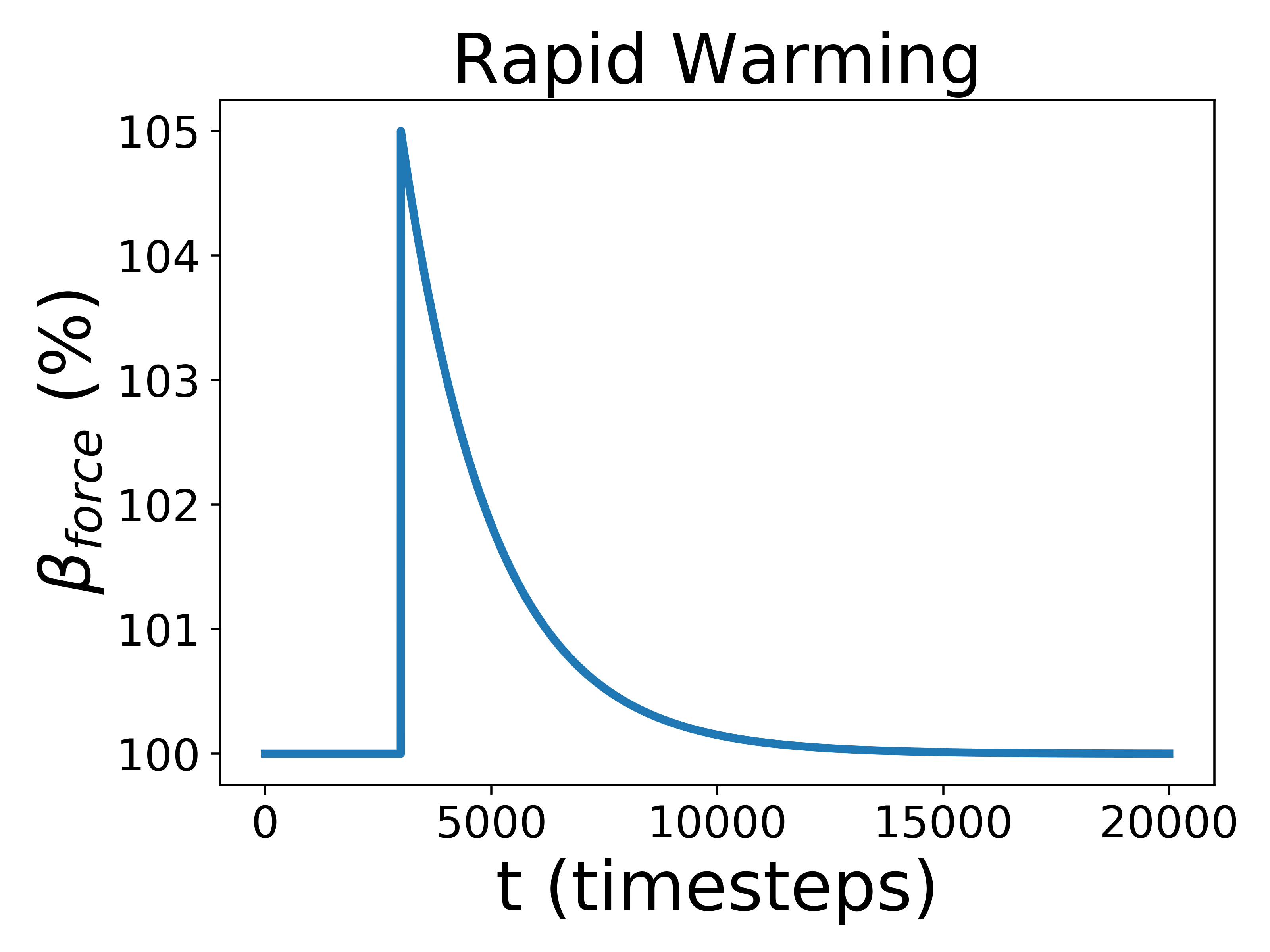}
	\label{fig:perturbation_hot_short_example}
	}
\sidesubfloat[]{
	\centering
	\includegraphics[width=0.4\columnwidth]{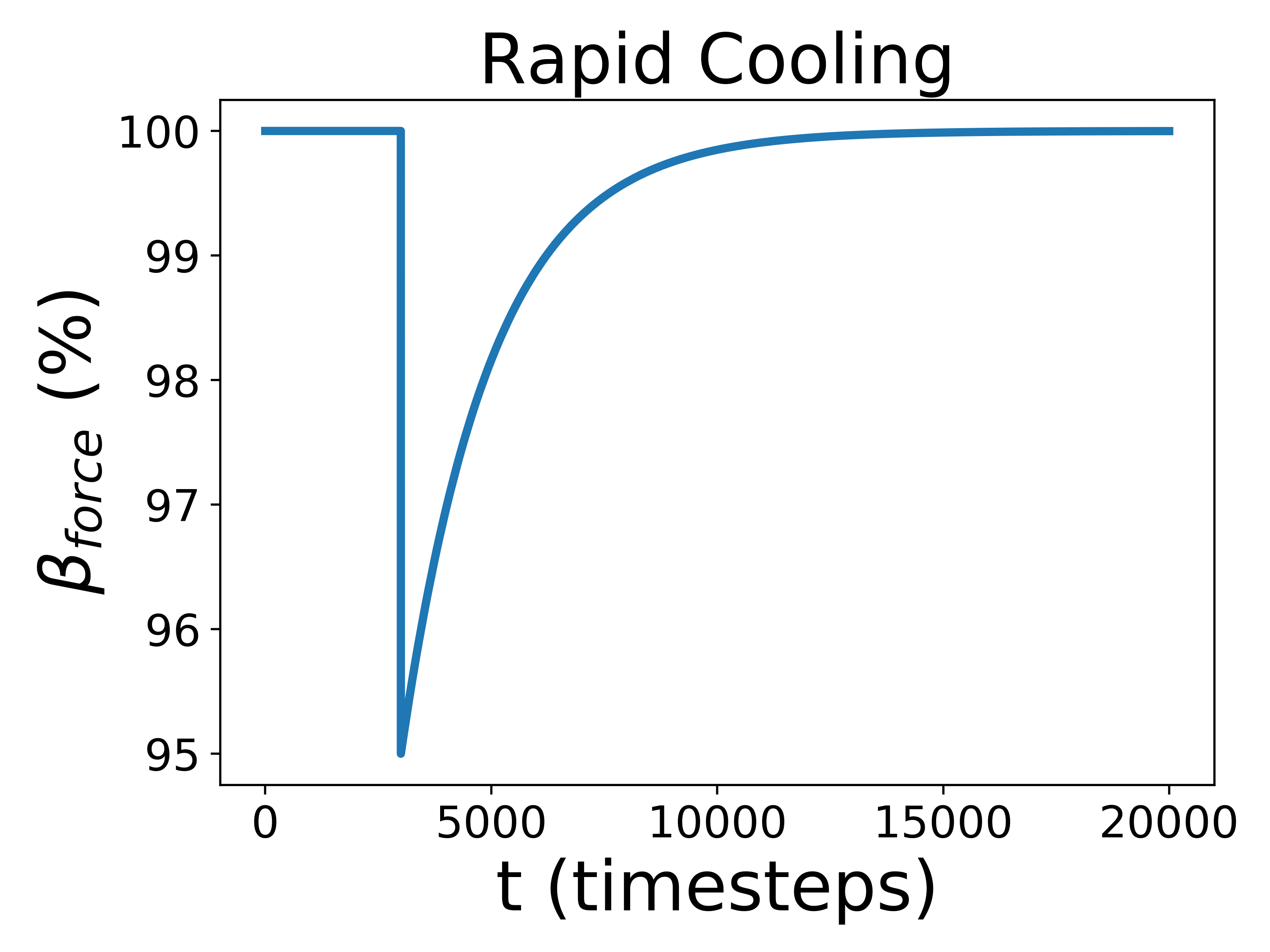}
	\label{fig:perturbation_cold_short_example}
	}
\caption{Changes in radiative forcing for different perturbations: constant radiative forcing (a), gradual warming (b), rapid warming (c), and rapid cooling (d).}
\label{fig:perturbation_examples}
\hfill
\end{figure}

To test how Gaian planets responded to changes in the radiative forcing, we introduced three different perturbations: one gradual and sustained linear increase, and two rapid perturbations, one positive and one negative, that increase or decrease abruptly before exponentially decaying back to the previous radiative forcing value.
The gradual scenario is analogous to stellar evolution; the rapid perturbation corresponds to impact events, volcanic eruptions, or snowball deglaciation.
For the rapid perturbation, we created a decay equation so that $\beta_{force}$ changed like so:
\par

\begin{equation} \label{eq:rapid_perturbation}
	\beta_{force} = \Delta\beta_{force}*exp(-(\frac{t-t_{pert}}{w}))
\end{equation}
where $\Delta\beta_{force}$ is the amount the radiative forcing increases by, $t$ is the current timestep, $t_{pert}$ is the timestep at which the perturbation occurs, and $w$ is the length of the perturbation.
In our experiment, $t_{pert}=3,000$ and $w=2,000$.
As for the gradual perturbations, $\beta_{force}$ just increases linearly:

\begin{equation}
\beta_{force}=t*(\frac{\beta_{force_f}-\beta_{force_o}}{t_{max}}) + \beta_{force_o}
\end{equation}
where $\beta_{force_f}$ is the final radiative forcing value, $\beta_{force_o}$ is the initial radiative forcing value, and $t_{max}$ is the total amount of timesteps. For use below, we define $\Delta\beta_{force}=\beta_{force_f}-\beta_{force_o}.$ All perturbations are demonstrated graphically in Figure \ref{fig:perturbation_examples}. \par

\floatsetup[figure]{style=plain,subcapbesideposition=top}

In our experiment, $\beta_{force_o} = 500$, consistent with \citet{nicholson2018gaian}.
For the rapid perturbation, we run scenarios in which the radiative forcing increases or decreases in intervals of $\pm 5\%$ of the initial forcing 
(in ExoGaia's radiative forcing units, $\pm 25$)
 until all of the planets become doomed, and for the longer evolution, we increase the radiative forcing by intervals of $\pm 30\%$, again until all the planets are doomed.
The extent to which this change in forcing alters the actual temperatures microbes experience depends on how well the microbes are able to regulate their host planet's environment (Eq. \ref{eq: climate}).
\par

\section{Results} \label{sec:results}

\floatsetup[figure]{style=plain,subcapbesideposition=top}
\begin{figure}
\centering
\sidesubfloat[]{
	\centering
	\includegraphics[width=0.45\columnwidth]{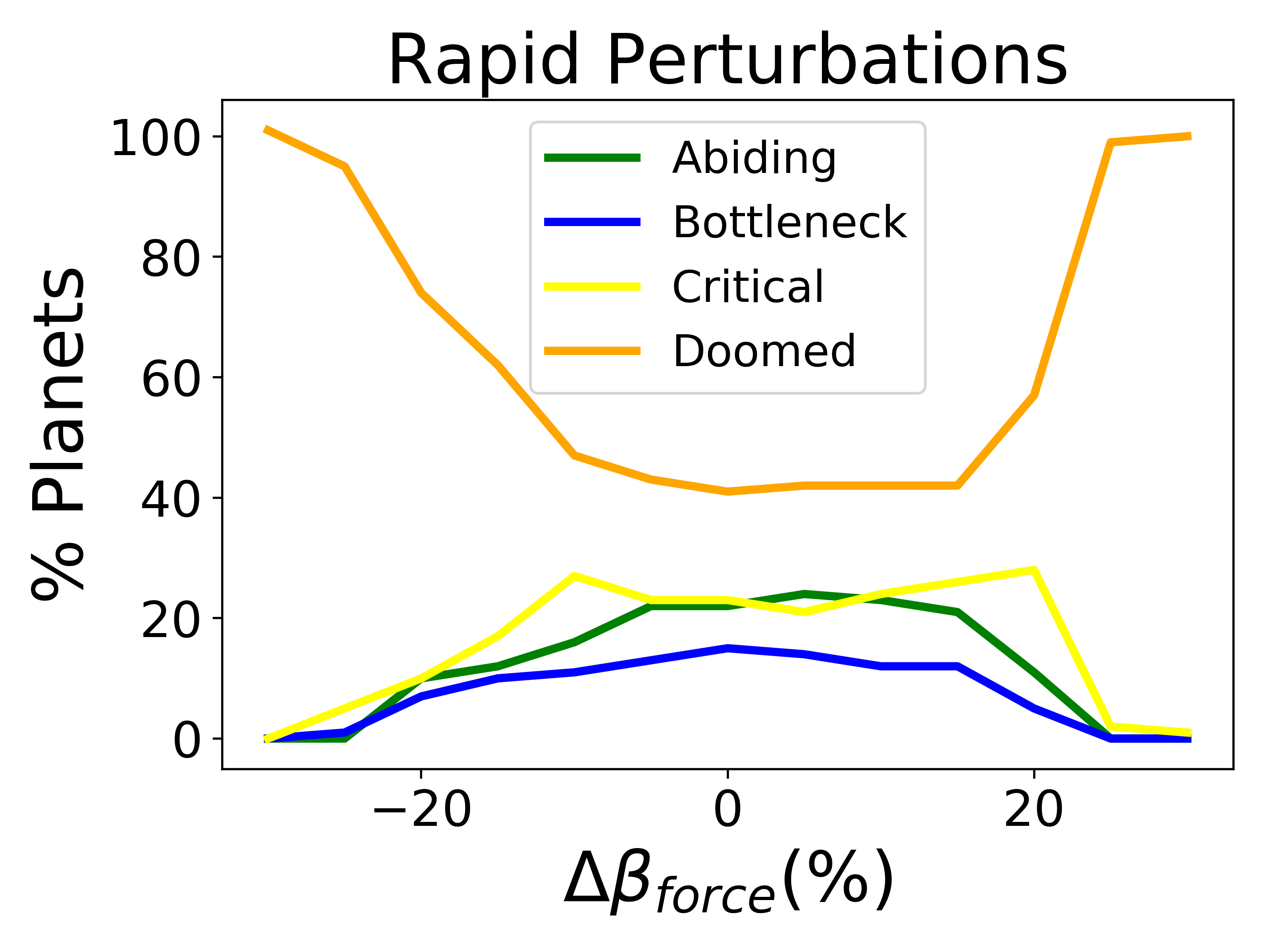}
	\label{fig:distributions_short}
	}
\sidesubfloat[]{
	\centering
	\includegraphics[width=0.45\columnwidth]{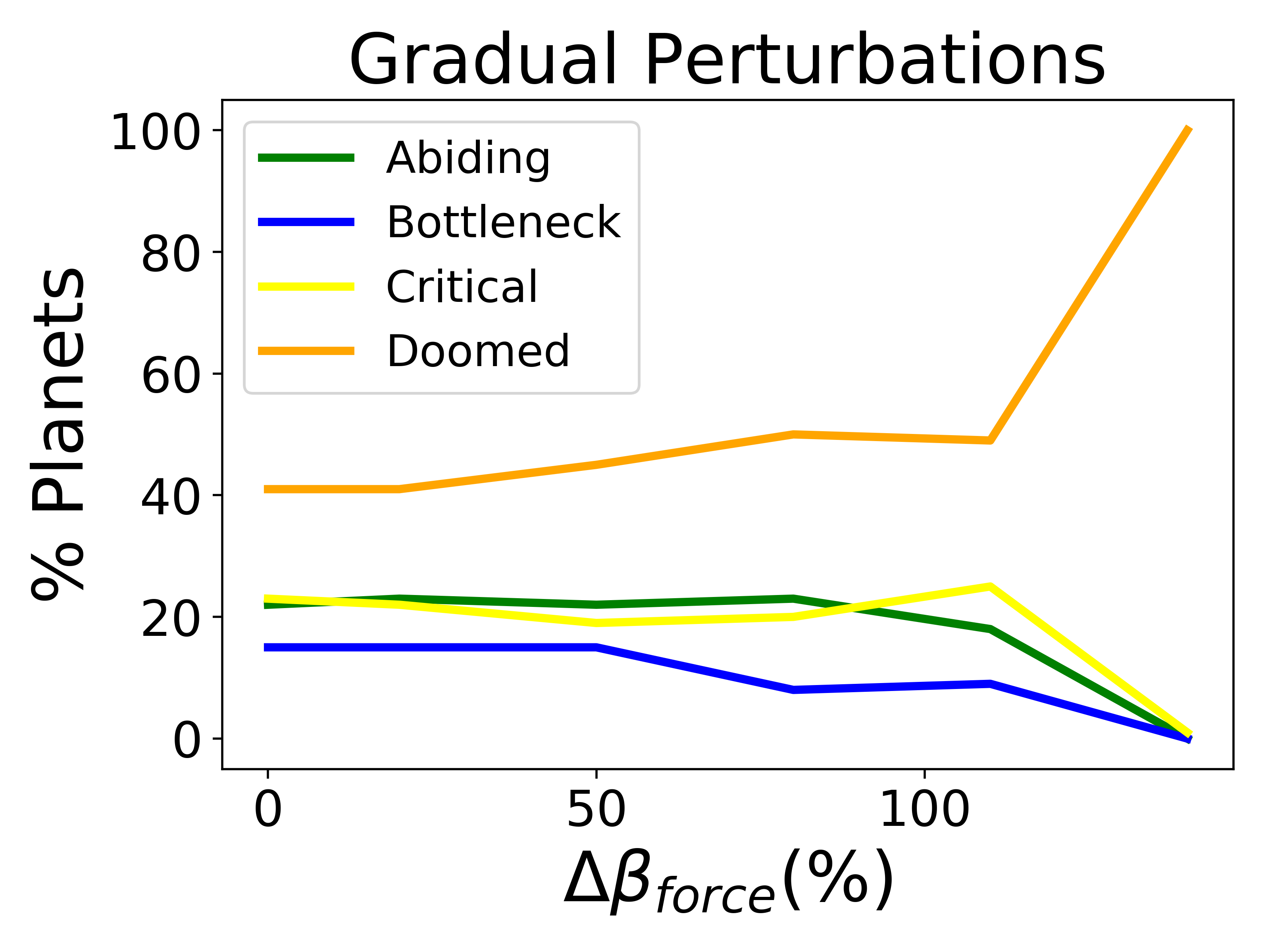}
	\label{fig:distributions_long}
	}
\caption{The distribution of planetary type for different perturbations. Figure (a) represents the distributions for rapid cooling and warming perturbations, and (b) for gradual warming. The green line represents the amount of abiding planets present for each perturbation, the blue line the amount of bottleneck, the yellow line critical, and the orange line doomed. $\Delta\beta_{force}$ represents the amount by which the radiative forcing is perturbed in terms of the percentage of the initial radiative forcing.}
\label{fig:distributions}
\end{figure}

\begin{figure}
\includegraphics[width=1\columnwidth]{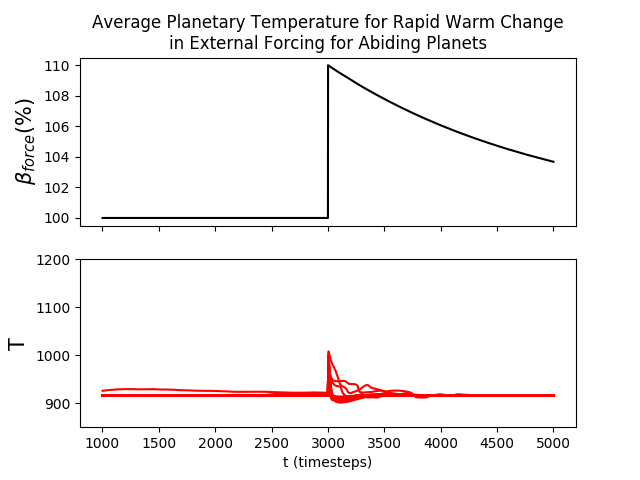}
\caption{The radiative forcing and the planetary temperature change for all 23 abiding planets experiencing the +10\% rapid perturbation. Each red line in the lower subplot represents the average planetary temperature across all 10 seedings of a planet in the time period immediately surrounding the perturbation.}
\label{fig:planetary_temp_example}
\end{figure}

Biospheres in ExoGaia can consistently survive rapid perturbations to the
initial radiative forcing
of about 15\%, and they have a similar response whether they are exposed to positive or negative perturbations (Fig. \ref{fig:distributions_short}).
The actual temperatures experienced by the biosphere differ with respect to how well the biosphere is able to regulate its environment. Well-regulated environments experience smaller and shorter changes in planetary temperatures than less well-regulated environments do.
Typically, when experiencing rapid warming perturbations, the planetary temperature on planets experiencing strong biological regulation briefly increases, but decreases back to the previous temperature very quickly (much faster than the perturbation decays) (Fig. \ref{fig:planetary_temp_example}). Planets with less robust regulations will undergo more abnormal temperature changes.
\par

\begin{figure}
\centering
\sidesubfloat[]{
	\includegraphics[width=1\columnwidth]{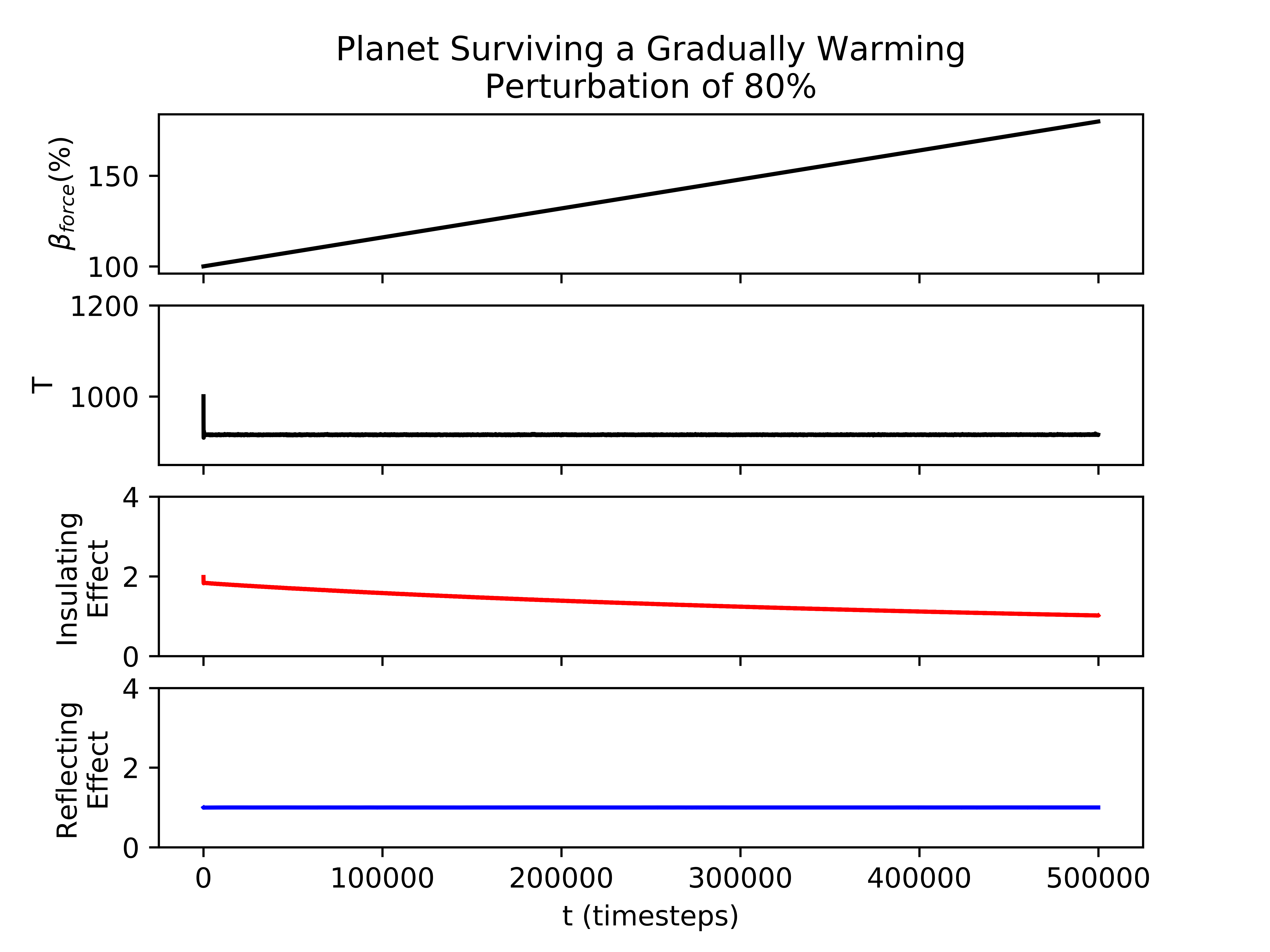}
	\label{fig:gradual_pert_planet60_80}
	} \\
\sidesubfloat[]{
	\includegraphics[width=1\columnwidth]{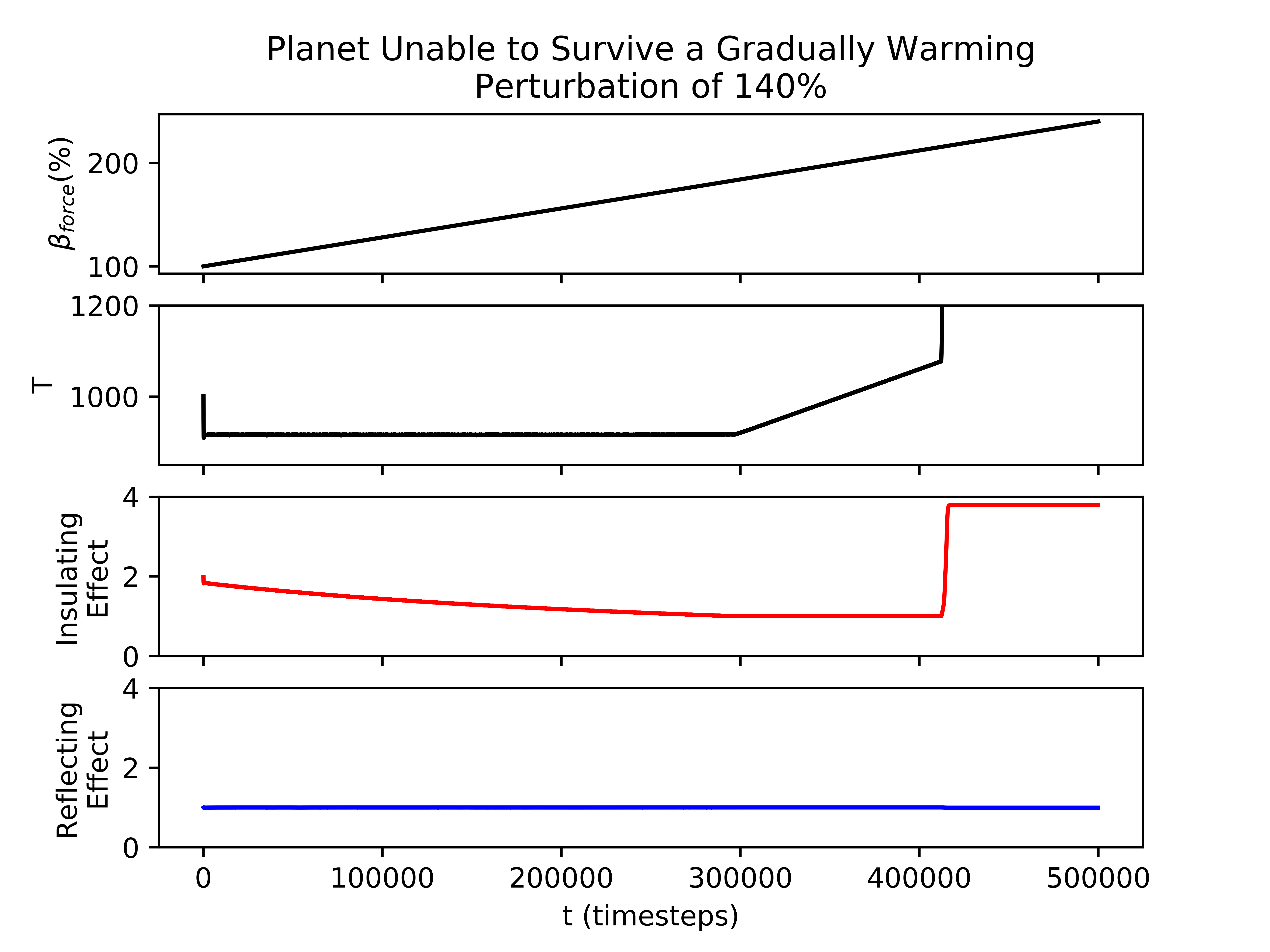}
	\label{fig:gradual_pert_planet60_140}
	}
\caption{An example of a planet experiencing gradual perturbations of +80\% (a) and +140\% (b) of the initial radiative forcing. The first subplot demonstrates the change to the radiative forcing and the second represents the average temperature across all 10 microbial seedings on this planet. The third subplot shows the average contribution across all 10 seedings that insulating chemicals make to the planetary temperature, given by $\frac{1}{1-AI}$, and the fourth shows the average contribution of reflecting chemicals to the planetary temperature, given by $1-AR$. AI is the fraction of the planet's thermal energy that is retained through insulation and AR is the fraction of incoming radiation reflected. The contributions made by each to the planetary temperature come from Equation \ref{eq: climate}.
}
\label{fig:gradual_planet60}
\end{figure}

Biospheres on planets experiencing gradual perturbations are much more robust than those experiencing rapid perturbations. A rapid change of about $\pm30\%$ of the initial radiative forcing ensures that all planets are doomed (Fig. \ref{fig:distributions_short}), but the radiative forcing must experience a 140\% increase from its initial value to achieve a similar result for the gradual scenario (Fig. \ref{fig:distributions_long}). 
Planets with strong biological regulation tend to have little to no fluctuation in planetary temperature in response to gradual perturbations (Fig. \ref{fig:gradual_pert_planet60_80}).
We define the insulating effect of the atmosphere on the planetary temperature as $\frac{1}{1-AI}$ and the reflecting effect of the atmosphere as $(1-AR)$, from Eq. \ref{eq: climate}.
Regulation occurs primarily from changes in the insulating effect of atmosphere, as opposed to reflecting (Fig. \ref{fig:gradual_pert_planet60_80}).
As the radiative forcing increases in Figure \ref{fig:gradual_pert_planet60_80}, the microbes reduce the abundances of insulating chemicals, and are able to regulate the environment enough to continue to survive. However, when the same planet experiences a larger gradual perturbation as in Figure \ref{fig:gradual_pert_planet60_140}, the abundances of insulating chemicals decrease and eventually reach zero, at which point the insulating effect of the atmosphere becomes 1 and the microbes are no longer able to regulate their environment. Once this occurs, the microbes become extinct and the planetary temperature rapidly increases to uninhabitable levels. Reflecting chemicals do not participate in the regulation of the planetary temperature, and the microbes are unable to increase the amount of reflecting chemicals as much as would be necessary to account for the increasing radiative forcing.
\par

\begin{figure}
\centering
\sidesubfloat[]{
	\centering
	\includegraphics[width=1\columnwidth]{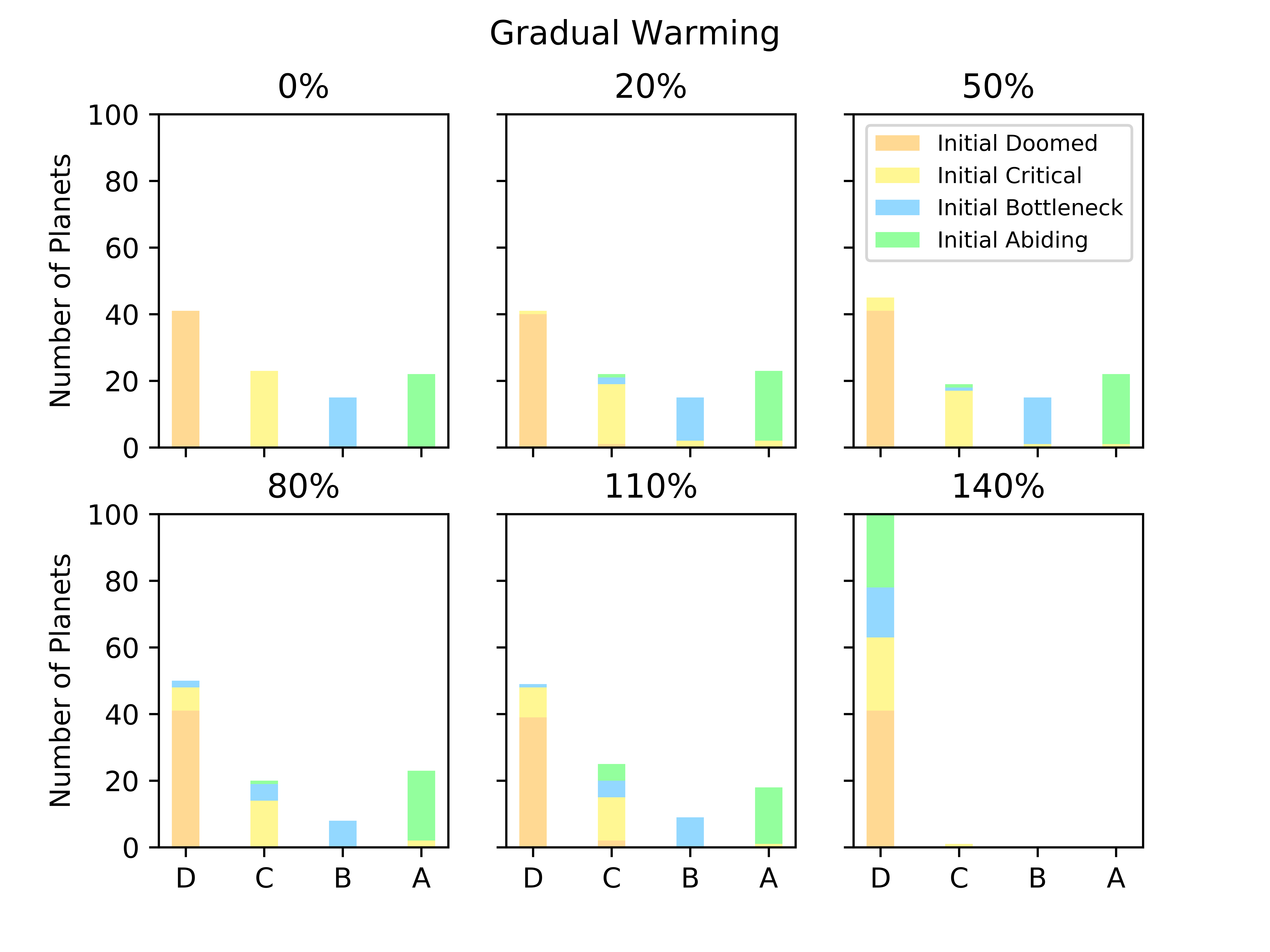}
	\label{fig:comparisons_subplots_long}
	} \\
\sidesubfloat[]{
	\centering
	\includegraphics[width=1\columnwidth]{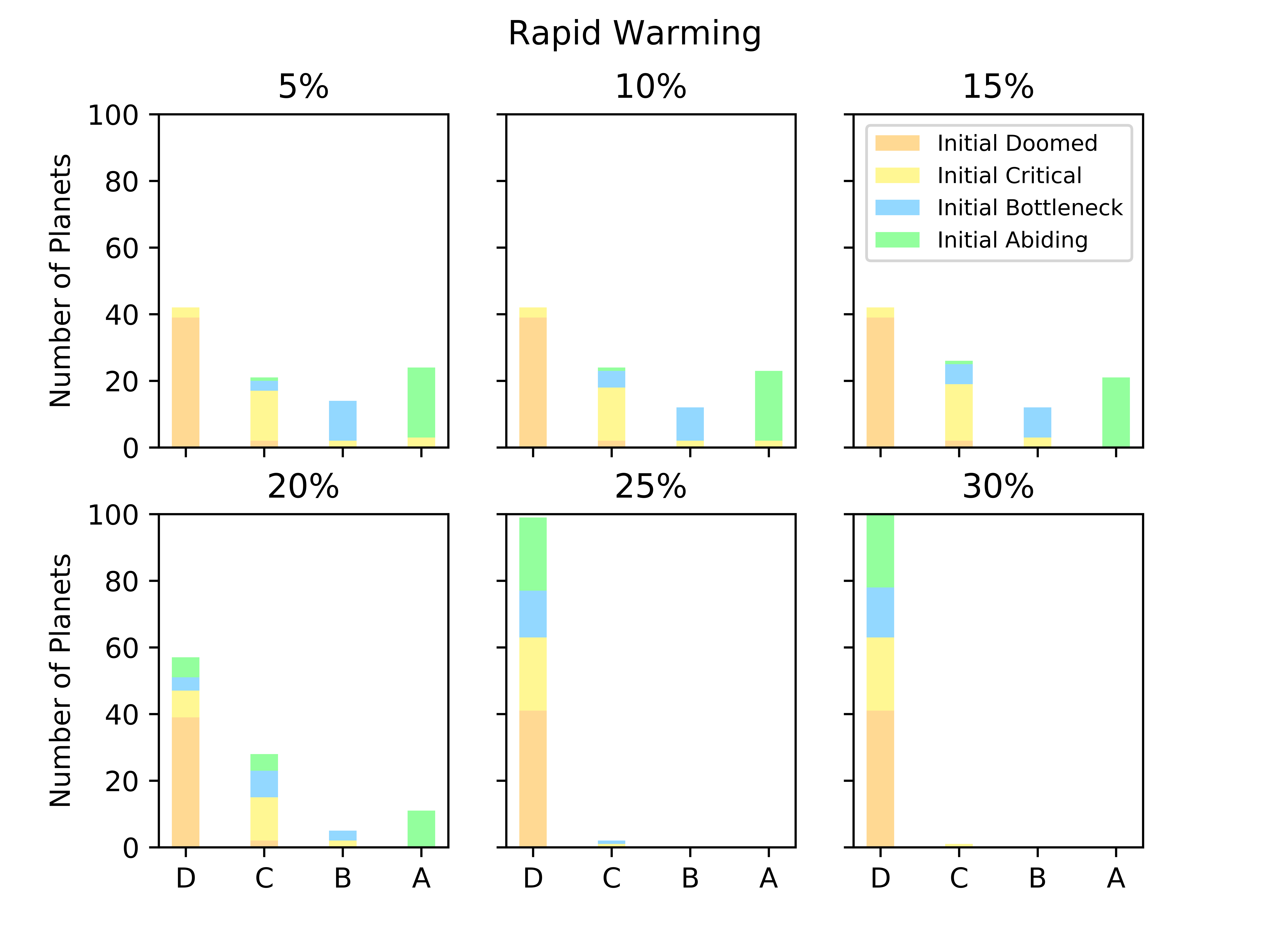}
	\label{fig:comparisons_subplots_short_hot}
	} \\
\sidesubfloat[]{
	\centering
	\includegraphics[width=1\columnwidth]{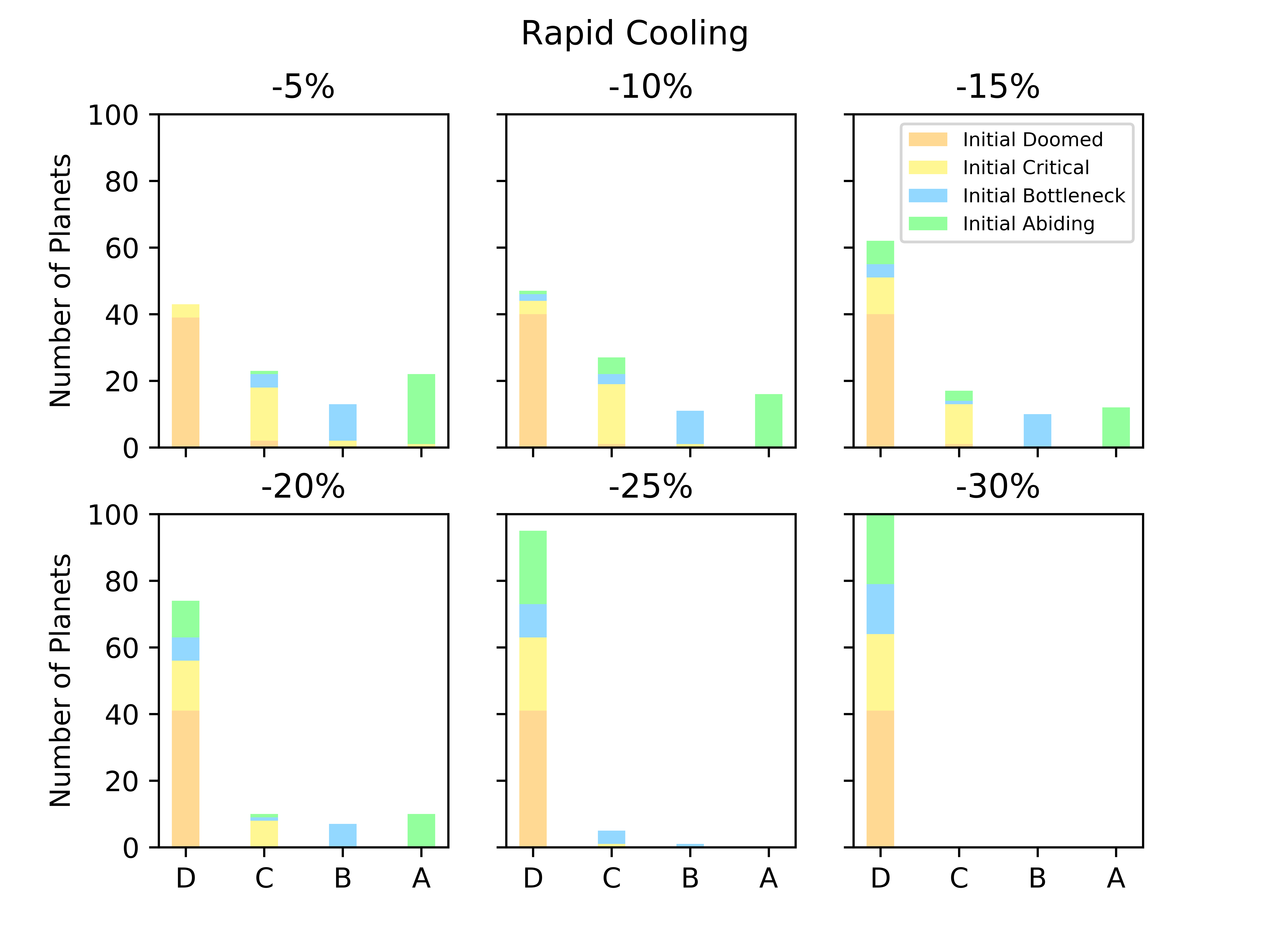}
	\label{fig:comparisons_subplots_short_cold}
	}
\caption{These plots depict how planetary type is distributed during different perturbations, in comparison to planetary type with no perturbation. All planets that were abiding with no perturbation are depicted in green, all that were bottleneck are depicted in blue, critical in yellow, and doomed in orange. The percentage above each subfigure is $\Delta\beta_{force}$, referring to the amount by which the radiative forcing changes from its initial value during a perturbation. D = Doomed, C = Critical, B = Bottleneck, and A = Abiding.}
\label{fig:comparisons_subplots}
\end{figure}

Gaian feedbacks are not inevitable and only emerge on a subset of planets even in the absence of external perturbation.
Most doomed planets stay doomed in the face of a perturbation, although a small number become critical. Planets that are abiding with constant radiative forcing are generally the same that are abiding with modest perturbations
(Fig. \ref{fig:comparisons_subplots}). There is more exchange between bottleneck and critical planets.

\floatsetup[figure]{style=plain,subcapbesideposition=top}

\floatsetup[figure]{style=plain,subcapbesideposition=top}

\begin{figure}
\includegraphics[width=1.0\columnwidth]{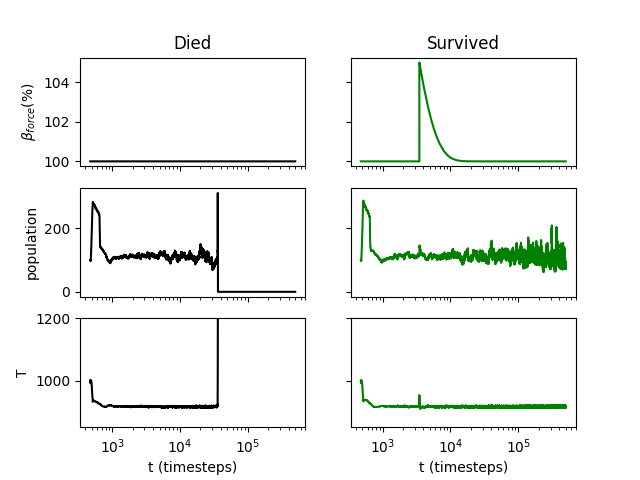}
\caption{
An example of a seeding on a planet that survives with a perturbation imposed, but dies without one.
The left column depicts the properties of the planet and life with no perturbation, and the right column depicts the same system but with a rapid, warm perturbation.}
\label{fig:planetcase}
\end{figure}

Interestingly, a few seedings on some planets
are actually more likely to survive with perturbations than without them (Fig. \ref{fig:comparisons_subplots}). Consider the example shown in Fig. \ref{fig:planetcase}.
When the planet does not experience a perturbation, this particular seeding of life
dies out during the latter half of the experiment, after experiencing large population and temperature spikes at approximately the 400,000th timestep.
However, when the planet experiences a perturbation in which the radiative forcing is briefly 5\% greater than it was initially, this seeding of life survives for the duration of the experiment. With the perturbation, although there are population spikes, there are no associated temperature spikes as dramatic as those with no perturbation.
\par
We do not have a definitive explanation for the small number of cases of seedings in which a perturbation improves survivability. It may be that global sterilization events require the confluence of a number of factors, and radiative perturbations can interfere with these confluences. In some cases, this can lead the biosphere to survive for the duration of the experiment, whereas without a perturbation it would have been unable to.
\par

\section{Discussion} \label{sec:discussion}

The qualitative results from this study provide intuition about the survival of life through perturbations to planetary systems.
Our results are particularly meaningful when considering the durability and longevity of biospheres on planets with Gaian feedbacks.
Because biospheres in ExoGaia can consistently survive
low-magnitude perturbations,
 it is unlikely that a small, abrupt warming or cooling on such a planet would be the sole reason that life on that planet would 
become extinct. Similarly, the
long-term evolution of a host star would be unlikely to kill off life on a planet unless it proceeded enough to cause a runaway greenhouse process.
\par
We have demonstrated that resilient biospheres are able to keep planetary environments stable even when they are subject to large, slow increases in the radiative forcing. This is particularly interesting given that regulation in ExoGaia only occurs because of biotic processes, and the model does not account for
Earth's abiotic carbon silicate cycle~\citep{Walker:1981aa}.
One implication of our results is that
 worlds lacking a functional carbonate-silicate cycle such as water-rich planets may enjoy long-term habitability if they become inhabited~\citep{Abbot_2015,kite2018habitability}.
\par
Although ExoGaia accounts for species evolution, it does not include ecological innovations such as burrowing or insulation, which could make species more likely to survive warmer or colder environments. None of the species are carnivorous, so there is no developed food chain, which could also impact the biosphere's regulatory ability. The zero-dimensionality of planets in ExoGaia prevents species from migrating during perturbations as they might on real planets, which could be a strategy for mitigating the most harmful effects from the perturbations on microbial populations. 
Planets in ExoGaia also do not have seasonality, which ~\citet{BITON2012145} found could decrease the range of luminosity values in which life is able to regulate the planetary environment, possibly making life less likely to survive larger scale perturbations.
\par
 The perturbations we consider in this study alter the radiation a planet is exposed to; however, we do not consider the effects of perturbations to chemical fluxes.
We also do not consider interacting perturbations, such as two impact events occurring simultaneously or soon after each other.
Both of these options can be implemented in the ExoGaia model, and are possibilities for future work.
\par
Perhaps the most important limitation of ExoGaia is its abstracted timescale. It is unclear how much real time on a planet ExoGaia's timesteps correspond to, which makes comparisons between ExoGaia and real planets difficult.
However, when implementing the rapid perturbations into our study, we found that changing the length of the perturbation ($w$ in Eq. \ref{eq:rapid_perturbation}) by two orders of magnitude (from 200 to 20,000 timesteps), did not qualitatively influence our results, so even though the timesteps are abstracted, it is likely that our qualitative findings remain correct.
\par
Our study demonstrates that Gaian climate regulation may be robust against perturbations. The effects of biotic regulation, both on Earth and on exoplanets, warrant further investigation.
Of particular interest will be determination of how Gaian feedbacks are impacted by increasing ecological complexity.
Additionally, \citet{nicholson2018gaian} found that higher probabilities of geochemical links tend to make planets more likely to have a regulating biosphere, and it would be worthwhile to continue to investigate characteristics that may contribute to a Gaian effect.
Finally, the effects of perturbations at different times with respect to the evolution of the biosphere, as opposed to simply testing perturbations early in the biosphere's evolution, remain to be explored.
\par

\section{Conclusions} \label{sec:conclusion}
By testing perturbations to the climate of planets using ExoGaia, we have arrived at the following findings:

\begin{enumerate}
	\item{The planets that are able to survive with constant radiative forcing are also those that are most likely to survive perturbations.}
	\item{Planets are able to tolerate gradual perturbations of significantly larger magnitude than rapid perturbations.}
	\item{The magnitude of change in radiative forcing matters more than whether the perturbations are warming or cooling.}
\end{enumerate}
These results imply that
life on other planets could survive the types of perturbations that Earth has experienced because of biological regulatory feedback loops.
Thus, if the Gaia hypothesis is correct, life is likely to be long-lived if it emerges on an exoplanet. Active biological regulation of atmospheric composition and the biospheric longevity that it imparts may ultimately improve our chances of finding life elsewhere in the Universe.
\par

\section*{Acknowledgements}

We thank Arwen Nicholson for sharing ExoGaia, and we thank Aaron L. Match and Qun Lu for their assistance implementing the code. We also acknowledge Thaddeus Komacek for providing feedback on an early draft. We thank the University of Chicago RCC for providing computer resources.
ODNA acknowledges support from the University of Chicago College
Research Fellows and President's Scholarship Programs. 
This work was supported by the NASA Astrobiology Program Grant Number
80NSSC18K0829 and benefited from participation in the NASA Nexus for
Exoplanet Systems Science research coordination network.
It was also partially supported by the National Science Foundation under NSF Grant
No. 1623064.
SLO acknowledges support from the T.C. Chamberlin postdoctoral fellowship in the Department of the Geophysical Sciences at the University of Chicago.




\bibliographystyle{mnras}
\bibliography{exogaia_references} 




\appendix


\bsp	
\label{lastpage}
\end{document}